\documentclass[letterpaper,journal]{IEEEtran}

\usepackage{amsmath,amsfonts,amssymb}
\usepackage{algorithm}
\usepackage{algorithmic}
\usepackage{array}
\usepackage{booktabs}
\usepackage{multirow}
\usepackage{cite}
\usepackage{graphicx}
\usepackage{textcomp}
\usepackage{tikz}
\usepackage{url}
\usepackage{xcolor}
\usetikzlibrary{arrows.meta,fit,positioning,shapes.geometric}

\floatname{algorithm}{Algorithm}

\makeatletter
\long\def\@makecaption#1#2{%
  \ifx\@captype\@IEEEtablestring
    \footnotesize\bgroup\par\centering
    \@IEEEtabletopskipstrut{\normalfont\footnotesize #1}\\%
    {\normalfont\footnotesize\scshape #2}\par
    \addvspace{0.5\baselineskip}\egroup
    \@IEEEtablecaptionsepspace
  \else
    \@IEEEfigurecaptionsepspace
    \parbox[t]{\hsize}{%
      \normalfont\footnotesize\centering
      {#1.}\nobreakspace\nobreakspace #2\par}%
  \fi}
\makeatother

\begin{document}
\raggedbottom

\title{FlowFPR: Flow Matching Prior for Oversampled Fourier Phase Retrieval}

\author{Jiatai~Li\textsuperscript{1},  Wenfei~Cao\textsuperscript{1,*},  Yang~Chen\textsuperscript{1}, and Kun~Miao\textsuperscript{2}%
\thanks{\textsuperscript{1}School of Mathematics and Statistics, Shaanxi Normal University, Xi'an 710119, China.}%
\thanks{\textsuperscript{2}Division of Data Information, Zhejiang National Mobilization Commission, Hangzhou 310002, China.}
\thanks{\textsuperscript{*}Corresponding author: Wenfei Cao (e-mail: caowenf2015@gmail.com).}}

\maketitle

\begin{abstract}
Fourier phase retrieval aims to reconstruct images from magnitude-only measurements; however, the absence of phase information, noises, and channel ambiguities make the recovery of natural color images challenging.
Although diffusion-prior-based methods impose strong image constraints, their recovery performance degrades under higher measurement noise, and repeated generative updates incur substantial computational overhead. To overcome these limitations, this paper proposes a novel method, termed FlowFPR, that integrates a pretrained flow-matching prior with the relaxed averaged alternating reflections (RAAR) algorithm for oversampled Fourier phase retrieval. In the basic version
of FlowPR  (Flow-RAAR-basic), a color image is estimated by repeatedly performing four steps at each outer iteration, where the flow-prior prediction step exploits single-step straight-line prediction, considerably reducing the generation cost. Nevertheless,  Flow-RAAR-basic still suffers from two critical issues: (i) the predefined fusion weights, \(\lambda_n\),  might lead to either ineffective or excessive correction, and (ii) the  fixed iteration number used by RAAR algorithm does not allow adaptation to measurement noise. To address these two issues, we develop two improved algorithms, Flow-RAAR-adaptive and Flow-RAAR-controlled, respectively. Moreover, these two improved algorithms naturally lead to a joint variant, Flow-RAAR-joint, which combines their advantages. Extensive experiments on a natural face image dataset demonstrate the superior performance of the four proposed methods compared with other relevant competing methods, as well as their acceleration over two recent diffusion-based methods. In the noise setting, Flow-RAAR-joint performs best among the four proposed methods.
\end{abstract}


\section{Introduction}
\label{sec:introduction}

Phase Retrieval (PR) reconstructs an unknown object from magnitude or intensity measurements without phase information.
PR is a fundamental inverse problem in coherent diffraction imaging, microscopy, crystallography, and astronomical imaging\cite{shechtman2015phase,miao1999noncrystalline}. Conventional optical detectors typically record intensity proportional to the squared field magnitude, whereas phase carries essential information about object morphology and spatial structure. Recovering phase from limited observations is therefore important for overcoming imaging-system limitations. Missing phase\cite{eldar2014phase, bendory2017non,vaswani2020nonconvex} also introduces nonconvexity, solution ambiguities, and noise sensitivity. For natural RGB images \cite{hayes1982reconstruction}, reconstruction must preserve both spatial structure within each channel and the color and geometric correspondence across channels, making recovery more challenging.

Classical iterative methods can be broadly organized by how their updates are constructed. One family develops projection, reflection, and feedback updates from measurement- and object-domain constraints, as exemplified by hybrid input--output (HIO) \cite{fienup1978modulus,fienup1982phase}, difference-map iterations \cite{elser2003projections}, and relaxed averaged alternating reflections (RAAR) \cite{luke2005raar}, with comparative evaluations of projection algorithms reported by Marchesini \cite{marchesini2007evaluation}. Another formulates an explicit optimization model and applies convex relaxation, alternating minimization, or nonconvex gradient iterations, as in PhaseLift \cite{candes2013phaselift}, alternating minimization with Gaussian measurements \cite{netrapalli2013alternating}, and Wirtinger Flow and its truncated variant \cite{candes2015wirtinger,chen2015truncated}. The former directly exploits Fourier magnitudes and known support, with early work establishing oversampled reconstruction of nonperiodic objects \cite{miao1998oversampling}; the latter provides recovery guarantees under specific measurement assumptions, which do not extend directly to arbitrary oversampled Fourier observations. Both remain limited by initialization, noise, and prior expressiveness, while channelwise reconstruction may yield equivalent solutions with inconsistent orientations \cite{li2024difffpr}.

Deep learning provides data-driven priors for phase retrieval, including supervised lensless imaging \cite{sinha2017lensless}, constraints imposed by the range of a generative network, initially studied for linear compressed sensing \cite{bora2017generative}, and untrained network architecture priors for image restoration \cite{ulyanov2018dip}. In an early study of phase retrieval with generative priors, Hand et al.\ analyze recovery properties under specified generator and measurement assumptions \cite{hand2018generative}. Plug-and-play methods \cite{venkatakrishnan2013pnp} instead embed learned denoisers, such as DnCNN \cite{zhang2017dncnn}, into measurement-driven iterations; DPIR demonstrates this modular design for image restoration \cite{zhang2022dpir}. For phase retrieval, prDeep combines a learned denoiser with regularization by denoising (RED) \cite{romano2017red,metzler2018prdeep}. These approaches separate the physical model from the prior but still require effective coordination of prior prediction and measurement correction.

Denoising diffusion and score-based models provide expressive generative priors \cite{ho2020ddpm,song2021sde}. Denoising Diffusion Restoration Models (DDRM) address linear inverse problems \cite{kawar2022ddrm}, whereas diffusion posterior sampling (DPS) supports noisy nonlinear phase retrieval through measurement-gradient guidance \cite{chung2023dps}. DiffFPR couples diffusion prediction with RAAR-type correction and candidate fusion for natural RGB images \cite{li2024difffpr}. DAPS, DDRM-PR, and I2I-PR further explore inverse-problem sampling with noise annealing, diffusion restoration with alternating projection, and learned reconstruction refinement, respectively \cite{zhang2025daps,kaya2025ddrmpr,kaya2025i2ipr}; their reported results are not directly comparable under our protocol. In our magnitude-domain Gaussian-noise experiments, DiffFPR's mean quality declines as noise increases, with failures on some images. Although DDIM offers a tradeoff between sampling quality and computational cost \cite{song2021ddim}, our multistep DiffFPR baseline remains costly under repeated restarts. These observations motivate improvements in noisy-reconstruction quality and computational efficiency.

Flow Matching learns a time-dependent velocity field connecting noise and data distributions, offering an alternative for efficient use of generative priors \cite{lipman2023flow}. PnP-Flow constructs a time-dependent image prior from a pretrained Flow model and demonstrates its applicability to image restoration \cite{martin2025pnpflow}. Recent work on FlowSGS combines split Gibbs sampling with Flow-based prior sampling and demonstrates nonlinear Fourier phase retrieval \cite{li2026flowsgs}. Unlike this posterior-sampling approach, our focus is on coupling a time-dependent Flow prediction with explicit RAAR correction and residual-guided fusion and inner search. Building on the time-dependent prior construction of PnP-Flow, we propose FlowFPR, which alternates a single Flow prediction with RAAR-type measurement correction and reconstructs the next path state after fusing the clean prediction and corrected candidate, without backpropagating measurement errors through the Flow network. Replacing the generative prior alone, however, does not ensure a suitable balance with measurement constraints. Fixed fusion weights or weights prescribed solely as a function of time can lead to ineffective or excessive data-consistency correction, while deeper inner searches can further fit measurement noise. We therefore introduce residual feedback in outer fusion and inner search, yielding Flow-RAAR-adaptive, Flow-RAAR-controlled, and their joint variant, alongside the basic method, Flow-RAAR-basic.

Our main contributions are as follows:
\begin{enumerate}
    \item We propose FlowFPR, combining a pretrained Flow Matching prior with RAAR-type measurement correction for oversampled Fourier phase retrieval of natural RGB images, with one time-dependent Flow prediction per outer iteration.
    \item We develop residual-improvement-based adaptive outer fusion and a controlled inner loop combining early stopping with best-so-far candidate selection. These yield two independent improvements and a joint variant that regulate the interaction between the generative prior and noisy measurements.
    \item We compare methods on 50 randomly selected FFHQ images at three noise levels and conduct ablations and runtime measurements on the same hardware on a 25-image subset. Under a common oracle best-of-20 protocol, the joint variant achieves the best mean PSNR, SSIM, and LPIPS at both noisy settings, while all four FlowFPR instances have lower measured end-to-end runtimes than DiffFPR.
\end{enumerate}

\section{Preliminaries}
\label{sec:preliminaries}

\subsection{Notation}
\label{subsec:notation}

We denote the ground-truth RGB image by $x^\star$, the state in the original image domain at outer iteration $n$ by $x_n$, and the oversampled Fourier-magnitude observation by $y$. The domains $\mathcal{X}$ and $\mathcal{U}_\kappa$ denote the original image domain and the oversampled domain, respectively. The symbols $H,W,C$ denote image height, width, and channel count, and $\kappa$ is the oversampling parameter. Domain dimensions and the measurement model are defined in Section~\ref{subsec:formulation}.

The subscript $n$ indexes outer iterations, and the superscript $(j)$ indexes inner candidates. The clean Flow prediction is $\hat{x}_n$, the $j$-th RAAR candidate is $x_n^{R(j)}$, and the candidate used for fusion is $x_n^R$. The actual number of inner iterations performed, $J_n$, must be distinguished from the best candidate index $j_n^\star$ among the candidates. The Flow time parameter is $t_n$, the stochastic path state in the normalized model domain is $z_{t_n}$, and the clean prediction in that domain is $\hat z_n$, distinct from the prediction $\hat{x}_n$ in the original image domain.

The operators $\mathcal{F}$ and $\mathcal{F}^{-1}$ denote channelwise two-dimensional Fourier transforms and their inverses; $\mathcal{P}_\kappa$ and $\mathcal{C}_\kappa$ denote centered zero padding and cropping; and $\mathcal{S}$ is the model-input normalization map. Elementwise interval clipping is $\Pi_{[a,b]}$, elementwise multiplication is $\odot$, and the Euclidean norm of a flattened array is $\|\cdot\|_2$. We use $[s]_+=\max(s,0)$; $I$ denotes an identity operator or matrix as appropriate. The Gaussian Flow-path sample $\xi$ and measurement noise $\eta$ have different roles, and $\varepsilon_{\mathrm{num}}>0$ protects against division by zero. The fusion weight $\lambda_n$, relative magnitude residual $r(x;y)$, and control parameters are defined in the subsequent section.

\subsection{Flow Matching Models}
\label{subsec:flow_prelim}

Flow Matching learns a time-dependent velocity field $v_\theta(z,t)$ that continuously transports a noise distribution to an image distribution \cite{lipman2023flow}. Its generative process $z(t)$ satisfies the following ordinary differential equation (ODE):
\begin{equation}
    \frac{\mathrm{d}z(t)}{\mathrm{d}t}=v_\theta(z(t),t),
    \qquad z(0)\sim\mathcal{N}(0,I),
\label{eq:flow_ode_prelim}
\end{equation}
where $t\in[0,1]$ and $\theta$ denotes network parameters. The endpoints $t=0$ and $t=1$ correspond to noise and data, respectively. Training directly regresses the target velocity field along a prescribed conditional probability path, thereby avoiding numerical integration of the full generative ODE at every training update.

We use a straight-path Flow model. For a normalized image sample $z_1$ and Gaussian sample $\xi$, the conditional path is $z_t=(1-t)\xi+tz_1$, with conditional target velocity $z_1-\xi$. This affine construction describes the training path; it does not imply that every generative trajectory of the learned velocity field is exactly straight. PnP-Flow uses this path and a pretrained velocity field to construct a time-dependent image prior \cite{martin2025pnpflow}. During reconstruction, we fix the model parameters and use a one-step clean prediction rather than solving the complete generative ODE at each correction. Normalization, stochastic path-state construction, and the implemented prior operator are defined in Section~\ref{subsec:basic}.


\subsection{RAAR Measurement-Consistency Correction}
\label{subsec:raar_prelim}

Let $y$ be an oversampled Fourier-magnitude observation and $u$ a spatial-domain variable of the same dimensions. The Fourier transform $\mathcal{F}$ and its inverse act independently on each color channel. The noisy-magnitude replacement, or data-consistency, operator used in the code is
\begin{equation}
    P_M^y(u)=\operatorname{Re}\!\left\{
    \mathcal{F}^{-1}\!\left[
    y\odot\frac{\mathcal{F}u}
    {\max\!\left(|\mathcal{F}u|,\varepsilon_{\mathrm{num}}\right)}
    \right]\right\},
\label{eq:magnitude_projection}
\end{equation}
where $\odot$ is elementwise multiplication and $\varepsilon_{\mathrm{num}}>0$ prevents division by zero in Fourier-phase normalization. Equation~\eqref{eq:magnitude_projection} preserves the current Fourier phase and replaces the magnitude with $y$. For nonnegative $y$ representing a consistent magnitude constraint, the operator can be interpreted as a projection onto the magnitude-constraint set. Under additive Gaussian noise, $y$ can have negative entries; $P_M^y$ then loses its strict set-projection interpretation and instead represents the noisy-magnitude replacement or data-consistency operator used by DiffFPR. We retain $P_M^y$ for consistency with common RAAR and DiffFPR notation.

Let $\Omega_\kappa$ denote the centered image-support region in the oversampled domain. The support projection $P_S$ is
\begin{equation}
    [P_S(u)]_p=
    \begin{cases}
        u_p, & p\in\Omega_\kappa,\\
        0,   & p\notin\Omega_\kappa.
    \end{cases}
\label{eq:support_projection}
\end{equation}
Using RAAR notation, define the magnitude-replacement reflector $R_M^y=2P_M^y-I$ and support reflector $R_S=2P_S-I$. The corresponding averaged-reflection term is
\begin{equation}
\begin{split}
    Q(u)&=\frac{1}{2}\left(R_SR_M^y+I\right)(u)\\
        &=P_S\!\left(2P_M^y(u)-u\right)+u-P_M^y(u).
\end{split}
\label{eq:q_operator}
\end{equation}
Luke's standard RAAR operator \cite{luke2005raar}, with relaxation parameter $\beta\in(0,1)$, is
\begin{equation}
    \mathcal{R}_\beta^{\mathrm{std}}(u;y)=
    \beta Q(u)+(1-\beta)P_M^y(u).
\label{eq:raar_standard}
\end{equation}
Li and Zhou also study RAAR for phase retrieval with structured illumination \cite{li2017raar}; their measurement setting differs from the oversampled Fourier-magnitude setting considered here.
Our experiments and the public DiffFPR implementation \cite{li2024difffpr} use a different operation order from \eqref{eq:raar_standard}: the final term reapplies $P_M^y$ to the averaged-reflection candidate $Q(u)$ rather than to the original input $u$. We refer to this code-consistent variant as \emph{post-projection relaxed RAAR}, where post-projection describes the additional data-consistency operation on $Q(u)$ in the second term of the relaxed combination:
\begin{equation}
    \mathcal{R}_\beta(u;y)=
    \beta Q(u)+(1-\beta)P_M^y\!\left(Q(u)\right).
\label{eq:raar_padded}
\end{equation}
This expression specifies the update used here and is distinct from Luke's standard RAAR. Unless otherwise stated, RAAR correction refers to the post-projection relaxed form in \eqref{eq:raar_padded}.

To connect the original image domain and the oversampled domain, let $\mathcal{P}_\kappa$ and $\mathcal{C}_\kappa$ denote centered zero padding and cropping, and $\Pi_{[0,1]}$ denote elementwise clipping. Domain dimensions are specified in Section~\ref{subsec:formulation}. Post-projection relaxed RAAR induces the following one-step composite correction in the original image domain:
\begin{equation}
    \mathcal{T}_\beta(x;y)=
    \Pi_{[0,1]}\!\left[
    \mathcal{C}_\kappa
    \mathcal{R}_\beta\!\left(\mathcal{P}_\kappa x;y\right)
    \right].
\label{eq:raar_image_operator}
\end{equation}
Given an initial estimate $\hat{x}_n$ in the original image domain, the dimensional transitions and updates of a fixed $K$-step correction are
\begin{align}
    x_n^{R(0)}&=\hat{x}_n\in\mathcal{X}, \notag \\
    u_n^{(j)}&=\mathcal{P}_\kappa x_n^{R(j-1)}\in\mathcal{U}_\kappa, \notag  \\
       v_n^{(j)}&=P_M^y\!\left(u_n^{(j)}\right)\in\mathcal{U}_\kappa,\notag \\
      w_n^{(j)}&=P_S\!\left(2v_n^{(j)}-u_n^{(j)}\right)
    +u_n^{(j)}-v_n^{(j)}\in\mathcal{U}_\kappa,\notag \\
    \bar{u}_n^{(j)}&=\beta w_n^{(j)}
    +(1-\beta)P_M^y\!\left(w_n^{(j)}\right)
    \in\mathcal{U}_\kappa, \notag \\
    x_n^{R(j)}&=\Pi_{[0,1]}\!\left(
    \mathcal{C}_\kappa\bar{u}_n^{(j)}\right)\in\mathcal{X},
    \quad j=1,\ldots,K.  \notag
\end{align}
Each inner iteration zero-pads the preceding candidate in the original image domain. Fourier transforms, magnitude replacement, support projection, and relaxation are performed in the oversampled domain with the same dimensions as $y$, followed by centered cropping and clipping. Fixed-step correction returns the $K$th candidate,
\begin{equation}
    x_n^R=x_n^{R(K)}=\mathcal{T}_\beta^K(\hat{x}_n;y).
\label{eq:fixed_raar}
\end{equation}

\section{Flow-Prior Phase Retrieval With Residual-Controlled RAAR}
\label{sec:method}
In this section, we propose a unified framework, termed FlowFPR, for oversampled Fourier phase retrieval of color images.
It combines a pretrained Flow Matching model as a natural-image prior with RAAR-type measurement-consistency correction, alternating between the original image domain and the oversampled measurement domain. The framework has four instances: Flow-RAAR-basic, Flow-RAAR-adaptive, Flow-RAAR-controlled, and their joint variant. Unlike approaches requiring multistep inversion along a complete generative trajectory, each instance makes only one time-dependent Flow prediction per outer iteration, followed by a small number of RAAR-type iterations to correct the Fourier magnitudes. They differ in their use of adaptive outer fusion and controlled inner search.

\begin{figure*}[t]
\centering
\includegraphics[width=0.94\textwidth]{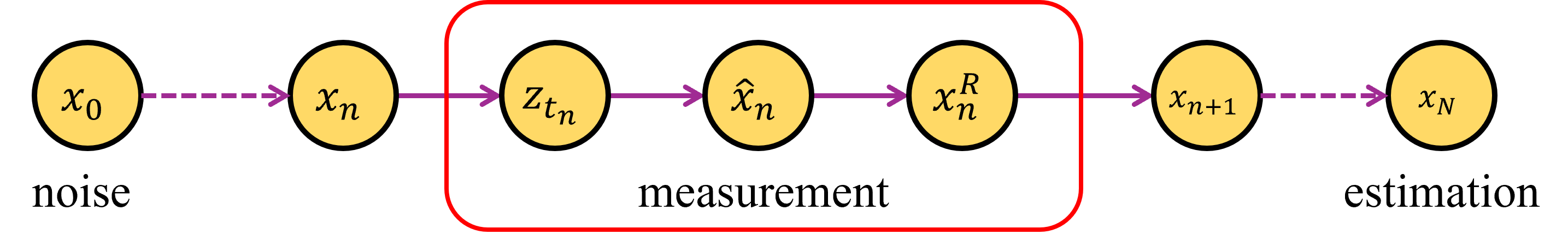}
\caption{FlowFPR reconstruction trajectory from the random initial state $x_0$ to the estimate $x_N$. Each outer iteration re-noises $x_n$ to obtain $z_{t_n}$, predicts $\hat{x}_n$ using the Flow prior, generates a RAAR candidate $x_n^R$, and fuses it with $\hat{x}_n$ to obtain $x_{n+1}$.}
\label{fig:overall_pipeline}
\end{figure*}
\subsection{Problem Formulation and Overall Framework}
\label{subsec:formulation}

Let the unknown color image be
\begin{equation}
    x^\star\in\mathcal{X}=[0,1]^{H\times W\times C},
\end{equation}
where $H$, $W$, and $C$ are image height, width, and channel count. Let $\kappa$ be the oversampling parameter used in the implementation and $p_\kappa$ the zero-padding width on each side. For the square images considered here, the implemented dimensions satisfy
\begin{equation}
    p_\kappa=\left\lfloor\frac{\kappa H}{8}\right\rfloor,\quad
 H_\kappa=H+2p_\kappa,\quad
    W_\kappa=W+2p_\kappa .
\label{eq:oversampled_size}
\end{equation}
Define centered zero padding $\mathcal{P}_\kappa:\mathcal{X}\rightarrow\mathcal{U}_\kappa$ and centered cropping $\mathcal{C}_\kappa:\mathcal{U}_\kappa\rightarrow\mathcal{X}$, where
\begin{equation}
    \mathcal{U}_\kappa=\mathbb{R}^{H_\kappa\times W_\kappa\times C}
\end{equation}
is the oversampled real-valued domain with the same dimensions as the observations. The operators $\mathcal{F}$ and $\mathcal{F}^{-1}$ are two-dimensional orthonormal discrete Fourier transforms and their inverses, acting independently on each color channel. The noiseless magnitude $b$ and observed measurement $y$ are
\begin{equation}
\begin{aligned}
    b=\left|\mathcal{F}\mathcal{P}_\kappa x^\star\right|,\quad   y=b+\eta, \quad \eta\sim\mathcal{N}(0,\sigma^2I).
\end{aligned}
\label{eq:measurement_model}
\end{equation}
where $b,y\in\mathcal{U}_\kappa$ and $\eta$ is additive magnitude-domain Gaussian noise. In particular, $y$ has the oversampled dimensions $H_\kappa\times W_\kappa\times C$, rather than the original image dimensions. For $H=W=256$ and $\kappa=4$, $p_\kappa=128$ and the observation is $512\times512\times3$, with four times as many measurement locations as original image pixels.

The proposed four methods alternate between two domains: Flow prediction, candidate fusion, and the outer state reside in the original image domain $\mathcal{X}$, while Fourier-magnitude replacement, support projection, and RAAR-type correction operate in the oversampled domain $\mathcal{U}_\kappa$. As shown in Fig.~\ref{fig:overall_pipeline}, 
at outer iteration $n$, the current clean state $x_n$ is re-noised at time $t_n$ to form $z_{t_n}$, and the velocity field produces the clean prior prediction $\hat{x}_n$. This prediction is zero-padded and corrected by RAAR to obtain the data-consistency candidate $x_n^R$. Finally, $\hat{x}_n$ and $x_n^R$ are fused in the original image domain to obtain the next clean state $x_{n+1}$.

\subsection{Flow-RAAR-basic}
\label{subsec:basic}
\subsubsection{Flow Prior Prediction}
Flow Matching learns a time-dependent velocity field $v_\theta(z,t)$ connecting a noise distribution and the natural-image distribution \cite{lipman2023flow}. To match the pretrained model's input range, define the normalization map and its inverse as
\begin{equation}
    \mathcal{S}(x)=2x-1,\qquad
    \mathcal{S}^{-1}(z)=\frac{z+1}{2}.
\label{eq:scale_map}
\end{equation}
For a state $x\in\mathcal{X}$ in the original image domain and a time parameter $t\in(0,1)$, the stochastic path state used in a Flow evaluation is
\begin{equation}
    z_t=t\mathcal{S}(x)+(1-t)\xi,\qquad
    \xi\sim\mathcal{N}(0,I).
\label{eq:flow_interpolation}
\end{equation}
Using the straight-path velocity prediction, a one-step clean-state estimate is
\begin{equation}
    \hat{z}=z_t+(1-t)v_\theta\!\left(z_t,t\right).
\label{eq:flow_clean_prediction}
\end{equation}
Clipping this prediction to its valid dynamic range gives the Flow prior operator used here:
\begin{equation}
    \mathcal{D}_\theta(x,t;\xi)=\mathcal{S}^{-1}\!\left[
    \Pi_{[-1,1]}\!\left(\hat{z}\right)\right].
\label{eq:flow_operator}
\end{equation}
where $\Pi_{[a,b]}$ is elementwise clipping to $[a,b]$. The Flow prior prediction at iteration $n$ is therefore
\begin{equation}
    \hat{x}_n=\mathcal{D}_\theta(x_n,t_n;\xi_n)\in\mathcal{X},
    \qquad \xi_n\sim\mathcal{N}(0,I).
\label{eq:flow_prior_step}
\end{equation}
This construction follows PnP-Flow's time-dependent generative-prior principle \cite{martin2025pnpflow}, while the subsequent data-consistency module is designed for oversampled Fourier-magnitude constraints.

\subsubsection{Fixed-Step RAAR Measurement-Consistency Correction}
The basic method directly applies the RAAR correction operator in the original image domain $\mathcal{T}_\beta$ defined in Section~\ref{subsec:raar_prelim}. Starting from the Flow prediction $\hat{x}_n$, it executes $K$ correction steps and uses the $K$th candidate $x_n^R$ in \eqref{eq:fixed_raar} for fusion, without introducing a new RAAR update rule.

\subsubsection{Damped Fusion}
The Flow prediction emphasizes the natural-image distribution, whereas the RAAR candidate emphasizes measurement consistency. Directly replacing $\hat{x}_n$ with $x_n^R$ can allow excessive data-consistency correction to disrupt the cross-channel structure provided by the generative prior. We therefore apply damped fusion in the original image domain:
\begin{equation}
    x_{n+1}=\Pi_{[0,1]}\!\left[
    \hat{x}_n+\lambda_n\left(x_n^R-\hat{x}_n\right)
    \right],
\label{eq:fusion_update}
\end{equation}
where $\lambda_n\in[\lambda_{\min},\lambda_{\max}]$ represents confidence in the RAAR correction direction at iteration $n$. The baseline fusion schedule $\lambda_n=g(t_n)$ should be chosen to match the generative prior and its probability path. Motivated by the signal coefficient $t$ in the straight path of \eqref{eq:flow_interpolation}, we adopt a linear baseline without additional shape parameters:
\begin{equation}
g(t)=\Pi_{[\lambda_{\min},\lambda_{\max}]}(t).
\label{eq:basic_lambda}
\end{equation}
It can be seen that the parameter setting, $\lambda_{\min}=0$ and $\lambda_{\max}=1$,  gives $\lambda_n=t_n$. 

\subsubsection{Time Evolution and  Re-noising}
The outer time parameter is discretized linearly:
\begin{equation}
    t_n=t_{\min}+\frac{n}{N-1}(t_{\max}-t_{\min}),
    \quad n=0,\ldots,N-1,
\label{eq:time_schedule}
\end{equation}
where $N\geq2$. In \eqref{eq:flow_interpolation}, smaller $t_n$ assigns a larger fraction to the random path component.
As $t_n$ increases, the path state retains more current-image information, while \eqref{eq:basic_lambda} progressively increases the RAAR correction weight. The fused image $x_{n+1}$ is the clean state for the next iteration in the original image domain. At iteration $n+1$, an independent $\xi_{n+1}$ is sampled and \eqref{eq:flow_interpolation} gives
\begin{equation}
\begin{aligned}
    z_{t_{n+1}}&=t_{n+1}\mathcal{S}(x_{n+1})
    +(1-t_{n+1})\xi_{n+1},\\
    \xi_{n+1}&\sim\mathcal{N}(0,I),
\end{aligned}
\label{eq:next_renoise}
\end{equation}
which maps the fused estimate to the next Flow time level. 

The outer loop starts at $t_{\min}$.  Accordingly,  image reconstruction starts from a state in the original image domain whose pixel values are independent and identically distributed uniform samples:
\begin{equation}
    [x_0]_{h,w,c}\overset{\mathrm{i.i.d.}}{\sim}\mathcal{U}(0,1).
\label{eq:random_init}
\end{equation}
The overall procedure proceeds through flow-prior prediction, measurement-consistency correction, damped fusion of the clean prediction with the RAAR-corrected result, and re-noising of the fused state, which then initializes the next iteration. The complete Flow-RAAR-basic procedure is summarized in Algorithm~\ref{alg:flow_raar_basic}.

However, the proposed Flow-RAAR-basic suffers from two critical issues:
\begin{itemize}
\item Issue 1: The predefined fusion weights \(\lambda_n\), which does not account for the residual information in the measurement correction through RAAR, can lead to either ineffective or excessive correction.
\item Issue 2: The RAAR algorithm employs a fixed iteration number \(K\), which does not allow adaptation to measurement noise.
\end{itemize}

\begin{algorithm}[t]
\caption{Flow-RAAR-basic Procedure}
\label{alg:flow_raar_basic}
\begin{algorithmic}[1]
\REQUIRE Magnitude measurement $y$, velocity field $v_\theta$, number of outer iterations $N\geq2$, fixed number of RAAR steps $K$, relaxation parameter $\beta$, time bounds $t_{\min},t_{\max}$, baseline schedule $g$, and weight bounds $[\lambda_{\min},\lambda_{\max}]$
\ENSURE Reconstructed image $x_N$
\STATE Independently sample $[x_0]_{h,w,c}\overset{\mathrm{i.i.d.}}{\sim}\mathcal{U}(0,1)$
\FOR{$n=0$ \TO $N-1$}
    \STATE $t_n\leftarrow t_{\min}+\dfrac{n}{N-1}(t_{\max}-t_{\min})$
    \STATE Independently sample $\xi_n\sim\mathcal{N}(0,I)$; $z_{t_n}\leftarrow t_n\mathcal{S}(x_n)+(1-t_n)\xi_n$
    \STATE $\hat z_n\leftarrow z_{t_n}+(1-t_n)v_\theta(z_{t_n},t_n)$, $\hat{x}_n\leftarrow\mathcal{S}^{-1}[\Pi_{[-1,1]}(\hat z_n)]$
    \STATE $x_n^{R(0)}\leftarrow\hat{x}_n$
    \FOR{$j=1$ \TO $K$}
        \STATE $x_n^{R(j)}\leftarrow\mathcal{T}_\beta(x_n^{R(j-1)};y)$
    \ENDFOR
    \STATE $\lambda_n\leftarrow g(t_n)$
    \STATE $x_{n+1}\leftarrow\Pi_{[0,1]}[\hat{x}_n+\lambda_n(x_n^{R(K)}-\hat{x}_n)]$
\ENDFOR
\RETURN $x_N$
\end{algorithmic}
\end{algorithm}
\subsection{Two Independent Residual-Guided Improvements}
\label{subsec:two_improvements}
Let us first define the relative magnitude residual of any state $x$ in the original image domain as follows:
\begin{equation}
    r(x;y)=
    \frac{\left\|\left|\mathcal{F}\mathcal{P}_\kappa x\right|-y\right\|_2}
    {\max\!\left(\|y\|_2,\varepsilon_{\mathrm{num}}\right)}.
\label{eq:relative_residual}
\end{equation}
Let $\hat{r}_n=r(\hat{x}_n;y)$ and $r_n^{R(j)}=r(x_n^{R(j)};y)$, and denote the final RAAR-candidate residual by $r_n^R=r(x_n^R;y)$. 

In the following, to address Issue 1, we propose an adaptive strategy for dynamically adjusting the fusion weights based on this residual, i.e., the Flow-RAAR-adaptive algorithm. Moreover, to address Issue 2, a best-candidate-solution scheme and an early-stopping scheme are adopted based on this residual to effectively control the iteration number $K$.
\subsubsection{Flow-RAAR-adaptive}
To quantitatively evaluate the residual improvement achieved by RAAR, we define the ratio of the post-correction residual to the pre-correction residual and its nonnegative variant as follows:
\begin{align}
    q_n=\frac{r_n^R}{\max(\hat{r}_n,\varepsilon_{\mathrm{num}})},~~d_n&=[1-q_n]_+,
    \label{eq:residual_improvement}
\end{align}
where $[s]_+=\max(s,0)$. For $q_n<1$, RAAR reduces the relative magnitude residual, and $d_n$ measures the relative improvement. For $q_n\geq1$, $d_n=0$, indicating that the candidate should not be fully trusted merely because of the time schedule.

Given a target improvement scale $\delta>0$ and factor bounds $[a_{\min},a_{\max}]$, define
\begin{align}
    a_n&=\Pi_{[a_{\min},a_{\max}]}\!\left(\frac{d_n}{\delta}\right),
    \label{eq:adaptive_factor}\\
    \lambda_n^{\mathrm{adp}}&=
    \Pi_{[\lambda_{\min},\lambda_{\max}]}\!\left[g(t_n)a_n\right].
    \label{eq:adaptive_lambda}
\end{align}
Flow-RAAR-adaptive still executes a RAAR loop with a fixed number of $K$ steps, with $x_n^R=x_n^{R(K)}$, but replaces the weight in \eqref{eq:fusion_update} by $\lambda_n^{\mathrm{adp}}$. We use $\delta=0.1$, $a_{\min}=0.2$, and $a_{\max}=1$, so the fusion weight adapts to the observed residual reduction within $[0.2g(t_n),g(t_n)]$ and never exceeds $g(t_n)$. For $q_n\geq1$, $d_n=0$ and $a_n=a_{\min}=0.2$. Thus, the clipping mechanism adaptively adjusts the RAAR candidate's fusion weight without completely rejecting the candidate. 

Overall, the proposed scheme above retains more of the Flow prediction when RAAR provides limited benefit, while maintaining a minimum RAAR correction strength and preventing the fusion weight from vanishing. 
Algorithm~\ref{alg:flow_raar_adaptive} gives the complete procedure. Relative to Algorithm~\ref{alg:flow_raar_basic}, the RAAR depth remains fixed at $K$, while the inputs additionally include $\delta$ and $[a_{\min},a_{\max}]$.
\begin{algorithm}[t]
\caption{Flow-RAAR-adaptive Procedure}
\label{alg:flow_raar_adaptive}
\begin{algorithmic}[1]
\REQUIRE Magnitude measurement $y$, velocity field $v_\theta$, number of outer iterations $N\geq2$, fixed number of RAAR steps $K$, relaxation parameter $\beta$, time bounds $t_{\min},t_{\max}$, baseline schedule $g$, weight bounds $[\lambda_{\min},\lambda_{\max}]$, improvement scale $\delta$, and factor bounds $[a_{\min},a_{\max}]$
\ENSURE Reconstructed image $x_N$
\STATE Independently sample $[x_0]_{h,w,c}\overset{\mathrm{i.i.d.}}{\sim}\mathcal{U}(0,1)$
\FOR{$n=0$ \TO $N-1$}
    \STATE $t_n\leftarrow t_{\min}+\dfrac{n}{N-1}(t_{\max}-t_{\min})$
    \STATE Independently sample $\xi_n\sim\mathcal{N}(0,I)$; $z_{t_n}\leftarrow t_n\mathcal{S}(x_n)+(1-t_n)\xi_n$
    \STATE $\hat z_n\leftarrow z_{t_n}+(1-t_n)v_\theta(z_{t_n},t_n)$, $\hat{x}_n\leftarrow\mathcal{S}^{-1}[\Pi_{[-1,1]}(\hat z_n)]$
    \STATE $x_n^{R(0)}\leftarrow\hat{x}_n$
    \FOR{$j=1$ \TO $K$}
        \STATE $x_n^{R(j)}\leftarrow\mathcal{T}_\beta(x_n^{R(j-1)};y)$
    \ENDFOR
    \STATE $\hat r_n\leftarrow r(\hat{x}_n;y)$, $r_n^R\leftarrow r(x_n^{R(K)};y)$
    \STATE $q_n\leftarrow r_n^R/\max(\hat r_n,\varepsilon_{\mathrm{num}})$
    \STATE $d_n\leftarrow[1-q_n]_+$, $a_n\leftarrow\Pi_{[a_{\min},a_{\max}]}(d_n/\delta)$
    \STATE $\lambda_n^{\mathrm{adp}}\leftarrow\Pi_{[\lambda_{\min},\lambda_{\max}]}[g(t_n)a_n]$
    \STATE $x_{n+1}\leftarrow\Pi_{[0,1]}[\hat{x}_n+\lambda_n^{\mathrm{adp}}(x_n^{R(K)}-\hat{x}_n)]$
\ENDFOR
\RETURN $x_N$
\end{algorithmic}
\end{algorithm}

\subsubsection{Flow-RAAR-controlled}
RAAR with a fixed inner iteration count always performs all updates and returns the 
$K$-th candidate. However, the depth required to achieve effective consistency varies across outer stages under noisy conditions. Flow-RAAR-controlled algorithm combines early stopping and best-so-far candidate selection to address this issue. 

In specific, given minimum and maximum numbers of inner iterations $K_{\min}$ and $K_{\max}$ and a target residual ratio $\rho\in(0,1)$, the actual inner stopping iteration number at outer iteration $n$ is
\begin{equation}
\begin{split}
    J_n=\min\Big(&\big\{j\in\{K_{\min},\ldots,K_{\max}\}\colon r_n^{R(j)} \leq \rho\hat r_n\big\}\\
    &\cup\{K_{\max}\}\Big).
\end{split}
\label{eq:controlled_stop}
\end{equation}
This quantity $J_n$ shows that the loop stops when the candidate residual reaches at most $\rho$ times the Flow-prediction residual at outer iteration $n$; if this target is not reached earlier, it runs to $K_{\max}$. Furthermore, to avoid returning a locally inferior candidate when RAAR terminates at $J_n$,  the best-so-far candidate is selected from all candidates generated up to termination:
\begin{equation}
    j_n^\star\in
    \mathop{\arg\min}_{0\leq j\leq J_n}r_n^{R(j)},
    \qquad x_n^R=x_n^{R(j_n^\star)}.
\label{eq:controlled_best}
\end{equation}
It is noted that including $j=0$ here allows $x_n^{R(0)}=\hat{x}_n$ to be returned if no RAAR update before stopping improves on the Flow prediction. Thus, $J_n$ specifies the number of inner iterations performed, while $j_n^\star$ identifies the candidate selected for fusion, with $0\leq j_n^\star\leq J_n\leq K_{\max}$. Outer fusion retains $\lambda_n=g(t_n)$. The main experiments use $K_{\min}=1$, $K_{\max}=10$, and $\rho=0.85$.

Algorithm~\ref{alg:flow_controlled_raar} describes the complete procedure. Early stopping within $[K_{\min},K_{\max}]$ 
replaces a fixed computational depth, and the best-so-far candidate is tracked until termination, thereby enabling adaptive control of the inner iteration number under noisy conditions.
\begin{algorithm}[t]
\caption{Flow-RAAR-controlled With a Controlled Inner Loop}
\label{alg:flow_controlled_raar}
\begin{algorithmic}[1]
\REQUIRE Magnitude measurement $y$, velocity field $v_\theta$, number of outer iterations $N\geq2$, relaxation parameter $\beta$, time bounds $t_{\min},t_{\max}$, baseline schedule $g$, weight bounds $[\lambda_{\min},\lambda_{\max}]$, minimum and maximum inner iteration counts $K_{\min},K_{\max}$, and target ratio $\rho$
\ENSURE Reconstructed image $x_N$
\STATE Independently sample $[x_0]_{h,w,c}\overset{\mathrm{i.i.d.}}{\sim}\mathcal{U}(0,1)$
\FOR{$n=0$ \TO $N-1$}
    \STATE $t_n\leftarrow t_{\min}+\dfrac{n}{N-1}(t_{\max}-t_{\min})$
    \STATE Independently sample $\xi_n\sim\mathcal{N}(0,I)$; $z_{t_n}\leftarrow t_n\mathcal{S}(x_n)+(1-t_n)\xi_n$
    \STATE $\hat z_n\leftarrow z_{t_n}+(1-t_n)v_\theta(z_{t_n},t_n)$, $\hat{x}_n\leftarrow\mathcal{S}^{-1}[\Pi_{[-1,1]}(\hat z_n)]$
    \STATE $\hat r_n\leftarrow r(\hat{x}_n;y)$
    \STATE $x_n^{R(0)}\leftarrow\hat{x}_n$, $j_n^\star\leftarrow0$, $r_{n,\mathrm{best}}\leftarrow\hat r_n$, $J_n\leftarrow K_{\max}$
    \FOR{$j=1$ \TO $K_{\max}$}
        \STATE $x_n^{R(j)}\leftarrow\mathcal{T}_\beta(x_n^{R(j-1)};y)$
        \STATE $r_n^{R(j)}\leftarrow r(x_n^{R(j)};y)$
        \IF{$r_n^{R(j)}<r_{n,\mathrm{best}}$}
            \STATE $j_n^\star\leftarrow j$, $r_{n,\mathrm{best}}\leftarrow r_n^{R(j)}$
        \ENDIF
        \IF{$j\geq K_{\min}$ \AND $r_n^{R(j)}\leq\rho\hat r_n$}
            \STATE $J_n\leftarrow j$; \textbf{break}
        \ENDIF
    \ENDFOR
    \STATE $x_n^R\leftarrow x_n^{R(j_n^\star)}$, $\lambda_n\leftarrow g(t_n)$
    \STATE $x_{n+1}\leftarrow\Pi_{[0,1]}[\hat{x}_n+\lambda_n(x_n^R-\hat{x}_n)]$
\ENDFOR
\RETURN $x_N$
\end{algorithmic}
\end{algorithm}

\subsection{Joint Variant}
\label{subsec:joint}
Flow-RAAR-adaptive governs the influence strength of the selected candidate on the outer state, while Flow-RAAR-controlled determines the evaluation depth of the candidate trajectory and the candidate to be returned. Since these two improvements act at different levels, they can be combined directly.
The joint variant first applies early stopping in \eqref{eq:controlled_stop}, selects $x_n^{R(j_n^\star)}$ using \eqref{eq:controlled_best}, and computes
\begin{equation}
  q_n^{\mathrm{joint}}=
    \frac{r_n^{R(j_n^\star)}}
    {\max(\hat{r}_n,\varepsilon_{\mathrm{num}})}, \label{eq:joint_ratio}
\end{equation}
which is substituted into \eqref{eq:residual_improvement}--\eqref{eq:adaptive_lambda} to obtain $\lambda_n^{\mathrm{adp}}$. The joint update is
\begin{equation}
\begin{split}
    x_{n+1}=\Pi_{[0,1]}\Big[
    \hat{x}_n+\lambda_n^{\mathrm{adp}}
    \big(x_n^{R(j_n^\star)}-\hat{x}_n\big)
    \Big].
\end{split}
\label{eq:joint_update}
\end{equation}
This design establishes two levels of residual feedback: the inner controller selects a candidate using stepwise residuals, while the outer controller sets the fusion strength from its final improvement over the Flow prediction. 

Algorithm~\ref{alg:flow_raar_joint} summarizes the joint variant. Its inputs include both the inner-control parameters $K_{\min},K_{\max},\rho$ and adaptive-fusion parameters $\delta,a_{\min},a_{\max}$; a fixed inner iteration count and return index $K$ are no longer used. Again, $J_n$ records the actual number of inner iterations performed, while $j_n^\star$ identifies the best candidate generated up to that point. The adaptive weight is computed from this best candidate's residual, rather than that of the candidate at the stopping position $J_n$.
\begin{algorithm}[t]
\caption{Flow-RAAR-joint  with Two-Level Residual Control}
\label{alg:flow_raar_joint}
\begin{algorithmic}[1]
\REQUIRE Magnitude measurement $y$, velocity field $v_\theta$, number of outer iterations $N\geq2$, relaxation parameter $\beta$, time bounds $t_{\min},t_{\max}$, baseline schedule $g$, weight bounds $[\lambda_{\min},\lambda_{\max}]$, minimum and maximum inner iteration counts $K_{\min},K_{\max}$, target ratio $\rho$, improvement scale $\delta$, and factor bounds $[a_{\min},a_{\max}]$
\ENSURE Reconstructed image $x_N$
\STATE Independently sample $[x_0]_{h,w,c}\overset{\mathrm{i.i.d.}}{\sim}\mathcal{U}(0,1)$
\FOR{$n=0$ \TO $N-1$}
    \STATE Execute steps 3--17 of Algorithm~\ref{alg:flow_controlled_raar}
    \STATE Execute steps 11--14 of Algorithm~\ref{alg:flow_raar_adaptive}, replacing $x_n^{R(K)}$ and $r_n^R$ by $x_n^{R(j_n^\star)}$ and $r_{n,\mathrm{best}}$, respectively
\ENDFOR
\RETURN $x_N$
\end{algorithmic}
\end{algorithm}

\section{Experiments}
\label{sec:experiments}

We evaluate the Flow--RAAR methods on the FFHQ face dataset. We first compare quantitative reconstruction performance under noiseless and noisy Fourier-magnitude measurements, then present representative visual results, and finally examine the data-consistency operator, iteration counts, and controlled-inner-loop parameters. Unless stated otherwise, all methods use the same test images, oversampling scheme, and noise model.

\subsection{Experimental Setup}
\label{subsec:experimental_setup}

\subsubsection{Dataset and Measurement Generation}
We randomly select 50 distinct images from FFHQ \cite{karras2019style}; their complete file identifiers are listed in the separate appendix. All images are converted to RGB and resized to $256\times256$. With $\kappa=4$ in \eqref{eq:oversampled_size}, the padding width is 128 on each side and the Fourier-magnitude observation $y$ has dimensions $512\times512\times3$. We consider the three magnitude-domain Gaussian noise levels
\begin{equation}
    \sigma\in\{0,0.03,0.05\}
\end{equation}
For a given image and noise level, all methods use the same clean image, noise-model parameters, and measurement-generation rule. Each method's program generates noise independently, however, so the exact noise realization is not forcibly shared. No nonnegativity clipping is applied after noise addition, and noisy $y$ may therefore contain negative entries.

\subsubsection{Comparison Methods}
\label{subsec:comparison_methods}

We compare eight methods: 1) classical HIO without a generative prior; 2) standalone RAAR using the post-projection relaxed update introduced in the preliminaries; 3) DPS, which guides diffusion-based reconstruction with measurement-likelihood gradients \cite{chung2023dps}; 4) DiffFPR, which alternates diffusion-prior prediction and RAAR correction; 5) Flow-RAAR-basic with fixed RAAR depth and a linear fusion-weight schedule; 6) Flow-RAAR-adaptive, which adjusts fusion weights according to residual improvement; 7) Flow-RAAR-controlled, which uses early stopping and returns the lowest-residual candidate generated before stopping; and 8) the joint variant combining controlled inner search and adaptive fusion. The Flow--RAAR methods share the same Flow model, time interval, and restart count. They differ only in the inner stopping and selection rules and outer fusion mechanism described in Section~\ref{sec:method}.

\subsubsection{Evaluation Metrics}
We assess reconstruction quality using peak signal-to-noise ratio (PSNR), structural similarity (SSIM) \cite{wang2004ssim}, and learned perceptual image patch similarity (LPIPS) \cite{zhang2018lpips}. Higher PSNR and SSIM are preferable, whereas lower LPIPS indicates closer agreement of deep perceptual features with the ground truth. LPIPS uses the public version 0.1 implementation with AlexNet features and RGB inputs normalized to $[-1,1]$. 
Relative magnitude residual \eqref{eq:relative_residual} additionally measures data consistency. Because a lower residual need not imply better visual quality, PSNR, SSIM, and LPIPS are the primary quality metrics, while magnitude residuals support controller analysis.

We additionally assess robustness to random initialization using the single-run success rate. A reconstruction from one random initialization is considered successful if its evaluation PSNR is strictly greater than 20~dB. The mean success rate (MSR), averaged over all test images and random initializations, is defined as
\begin{equation}
    \mathrm{MSR}=\frac{1}{ML}\sum_{i=1}^{M}\sum_{\ell=1}^{L}
    \mathbb{I}\!\left(\mathrm{PSNR}_{i,\ell}>20~\mathrm{dB}\right)\times 100\%,
\label{eq:mean_success_rate}
\end{equation}
where $M=50$ is the number of test images, $L=20$ is the number of random initializations per image, $\mathbb{I}(\cdot)$ is the indicator function, and $\mathrm{PSNR}_{i,\ell}$ is the evaluation value for the $\ell$th independent run on image $i$. This metric is equivalent to computing the single-run success rate for each image and then averaging over the 50 images. Higher MSR indicates greater robustness to random initialization.

\subsubsection{Implementation Details}
The Flow--RAAR methods use the official CelebA-pretrained optimal-transport (OT) flow matching U-Net weights released with PnP-Flow \cite{martin2025pnpflow}; this model was not trained on FFHQ. The authors state that the public checkpoint results from subsequent retraining and is not identical to the checkpoint used in their paper. It was trained on CelebA at $128\times128$. Without retraining or fine-tuning, we exploit the U-Net's fully convolutional structure to apply it directly to $256\times256$ FFHQ images. We use $N=800$ outer iterations, times from $t_{\min}=0.25$ to $t_{\max}=0.95$, and RAAR relaxation $\beta=0.75$. Flow-RAAR-basic and Flow-RAAR-adaptive use fixed inner depth $K=10$; Flow-RAAR-controlled and the joint variant use early stopping and best-so-far selection with $K_{\min}=1$, $K_{\max}=10$, and $\rho=0.85$. Baseline fusion uses $g(t)=t$ and $[\lambda_{\min},\lambda_{\max}]=[0,1]$, while adaptive fusion uses $\delta=0.1$, $a_{\min}=0.2$, and $a_{\max}=1$. Table~\ref{tab:main_parameters} summarizes the principal parameters.

\begin{table}[t]
\caption{Main Experimental Parameters of the FlowPR Methods}
\label{tab:main_parameters}
\centering
\renewcommand{\arraystretch}{1.08}
\begin{tabular}{lc}
\toprule
Parameter & Value \\
\midrule
Image / observation size & $256\times256$ / $512\times512$ \\
Oversampling parameter $\kappa$ & 4 \\
Outer iterations $N$ & 800 \\
$t_{\min},t_{\max}$ & $0.25,0.95$ \\
Baseline fusion schedule $g(t)$ & $t$ \\
$\lambda_{\min},\lambda_{\max}$ & $0,1$ \\
RAAR relaxation $\beta$ & 0.75 \\
Fixed inner depth $K$ & 10 \\
Controlled loop $(K_{\min},K_{\max},\rho)$ & $(1,10,0.85)$ \\
$\delta,a_{\min},a_{\max}$ & $0.1,0.2,1.0$ \\
Random restarts & 20 \\
\bottomrule
\end{tabular}
\end{table}

To reduce sensitivity to random initialization in this nonconvex problem, each image is reconstructed independently 20 times
and  the reconstructed image with the highest PSNR is exploited as the final result.
DPS uses the public unconditional FFHQ diffusion weights, 1000-step DDPM sampling, and the default posterior-correction scale of 1.0 \cite{chung2023dps}. To match our measurements, its clean image prediction in $[-1,1]$ is mapped to $[0,1]$ before applying the fourfold-oversampled Fourier-magnitude operator in the data-consistency gradient. DiffFPR uses its FFHQ diffusion model and original 1000-step setting \cite{li2024difffpr}. Standalone HIO uses random-phase initialization, 1000 iterations, and $\beta_{\mathrm{HIO}}=0.9$ \cite{fienup1982phase}. Standalone RAAR uses random Fourier-phase initialization, 1000 iterations, and $\beta=0.75$,  with the same post-projection relaxed update as the measurement-correction module. The implementation uses PyTorch 2.6.0 and CUDA 12.4. All runtime measurements are conducted on the same NVIDIA GeForce RTX 4070 SUPER GPU.

\subsection{Quantitative Comparison}
\label{subsec:quantitative_comparison}

Table~\ref{tab:ffhq25_main_comparison} reports mean PSNR, SSIM, and LPIPS over all 50 randomly selected FFHQ images. Each image contributes equally, and boldface identifies the best result at each noise level.

\begin{table*}[t]
\caption{Mean Reconstruction Results on 50 FFHQ Images at Different Noise Levels}
\label{tab:ffhq25_main_comparison}
\centering
\renewcommand{\arraystretch}{1.12}
\scriptsize
\setlength{\tabcolsep}{4.0pt}
\begin{tabular*}{\textwidth}{@{\extracolsep{\fill}}l*{12}{c}}
\toprule
& \multicolumn{4}{c}{$\sigma=0$} & \multicolumn{4}{c}{$\sigma=0.03$} & \multicolumn{4}{c}{$\sigma=0.05$} \\
\cmidrule(lr){2-5}\cmidrule(lr){6-9}\cmidrule(lr){10-13}
Method & PSNR & SSIM & LPIPS & MSR (\%) & PSNR & SSIM & LPIPS & MSR (\%) & PSNR & SSIM & LPIPS & MSR (\%) \\
\midrule
HIO & 21.3528 & 0.6409 & 0.4377 & 43.9 & 16.2323 & 0.2817 & 0.7540 & 2.3 & 14.7042 & 0.2011 & 0.8135 & 0.0 \\
RAAR & 18.5623 & 0.4349 & 0.6479 & 5.5 & 18.3807 & 0.3667 & 0.6976 & 4.9 & 18.2471 & 0.3232 & 0.7148 & 4.7 \\
DPS & 27.7673 & 0.7908 & 0.1333 & 42.4 & 24.9240 & 0.7104 & 0.1722 & 40.4 & 23.8213 & 0.6747 & 0.1898 & 39.9 \\
DiffFPR & 36.9518 & 0.9005 & 0.0779 & 68.4 & 25.3380 & 0.6032 & 0.3041 & 66.6 & 23.0191 & 0.4566 & 0.4589 & 65.0 \\
Flow-RAAR-basic & \textbf{47.1896} & \textbf{0.9932} & \textbf{0.0013} & 59.3 & 27.7261 & 0.6245 & 0.2447 & 55.2 & 24.3497 & 0.4558 & 0.4323 & 52.0 \\
Flow-RAAR-adaptive & 46.6506 & 0.9924 & 0.0017 & 59.1 & 30.3574 & 0.7644 & 0.1283 & 55.8 & 28.7971 & 0.7491 & 0.1605 & 56.0 \\
Flow-RAAR-controlled & 42.1632 & 0.9694 & 0.0129 & \textbf{75.7} & 28.8841 & 0.6774 & 0.1789 & 74.5 & 27.5820 & 0.6223 & 0.2169 & 71.7 \\
Flow-RAAR-joint & 42.5978 & 0.9734 & 0.0114 & \textbf{75.7} & \textbf{30.7169} & \textbf{0.7711} & \textbf{0.1225} & \textbf{74.7} & \textbf{30.4736} & \textbf{0.7989} & \textbf{0.1052} & \textbf{73.0} \\
\bottomrule
\end{tabular*}
\par\smallskip
\parbox{\textwidth}{\scriptsize\raggedright Note: DiffFPR uses the channelwise orientation rule in its public implementation; all other methods report unaligned RGB metrics. All results use ground-truth-assisted best-of-20 restart selection. MSR (mean success rate), defined in \eqref{eq:mean_success_rate}, uses a final evaluation PSNR strictly greater than 20~dB as the success criterion.}
\end{table*}

\textit{Role of learned priors.} Without measurement noise, all learning-based methods in Table~\ref{tab:ffhq25_main_comparison} outperform HIO and standalone RAAR in mean PSNR, SSIM, and LPIPS. Magnitude and support constraints cannot distinguish certain channelwise orientation combinations, and measurement compatibility alone does not ensure a natural RGB reconstruction. Generative priors supply facial structure, color, and cross-channel statistics, favoring structurally and chromatically coherent candidates beyond agreement with the observations.

\textit{Sources of Basic's advantage.} Flow-RAAR-basic outperforms DiffFPR and DPS on all three mean metrics without noise. DPS backpropagates measurement error through the generative network; its gradient update is not equivalent to directly satisfying magnitude and support constraints. Basic instead explicitly applies RAAR to a clean prediction. It performs $K=10$ composite corrections in the original image domain per outer iteration, compared with one oversampled-domain update in the DiffFPR implementation used here, making fuller use of exact measurements. The schedule $g(t_n)=t_n$ retains more generative structure early and strengthens measurement correction later, before the fused image undergoes further prior adjustment. The gain therefore concerns the combined pretrained prior, correction depth, and fusion schedule, rather than the Flow model alone.

\textit{Interaction under noisy measurements.} At $\sigma=0.03$ and $0.05$, all three improved variants outperform Basic on the three mean quality metrics; the joint variant increases PSNR by 2.9908 and 6.1239~dB, respectively. Consistency with noisy magnitudes does not coincide with ground-truth recovery, and deeper correction may further fit noise. The controlled inner loop truncates the candidate set upon reaching a relative residual target and returns the best-so-far candidate among generated candidates. Adaptive fusion instead adjusts correction strength according to residual improvement, retaining more prior information when the benefit is limited. These mechanisms regulate search depth and candidate influence, respectively, thereby adjusting the prior--measurement interaction. Early stopping applies only to the controlled and joint variants; residual feedback reflects correction benefit rather than directly identifying noise.

\subsection{Qualitative Results}
\label{subsec:qualitative_comparison}

The noiseless panel retains images 19123, 27175, 03899, and 04561. At each nonzero noise level, we randomly select two DiffFPR failures (PSNR below 20~dB) and two successful cases (PSNR at least 20~dB) from the 50-image test set, excluding images already used in another panel. The three panels therefore display 12 distinct images. The first two rows of each noisy panel show failures, and the last two show successful cases. This stratified sampling illustrates both outcomes, not their empirical frequencies. All panels use the same method-column order. Flow--RAAR, HIO, and DPS show original RGB reconstructions; DiffFPR follows the channelwise ambiguity-resolution rule in its public code, consistent with Table~\ref{tab:ffhq25_main_comparison}.

\begin{figure*}[t]
\centering
\includegraphics[width=\textwidth]{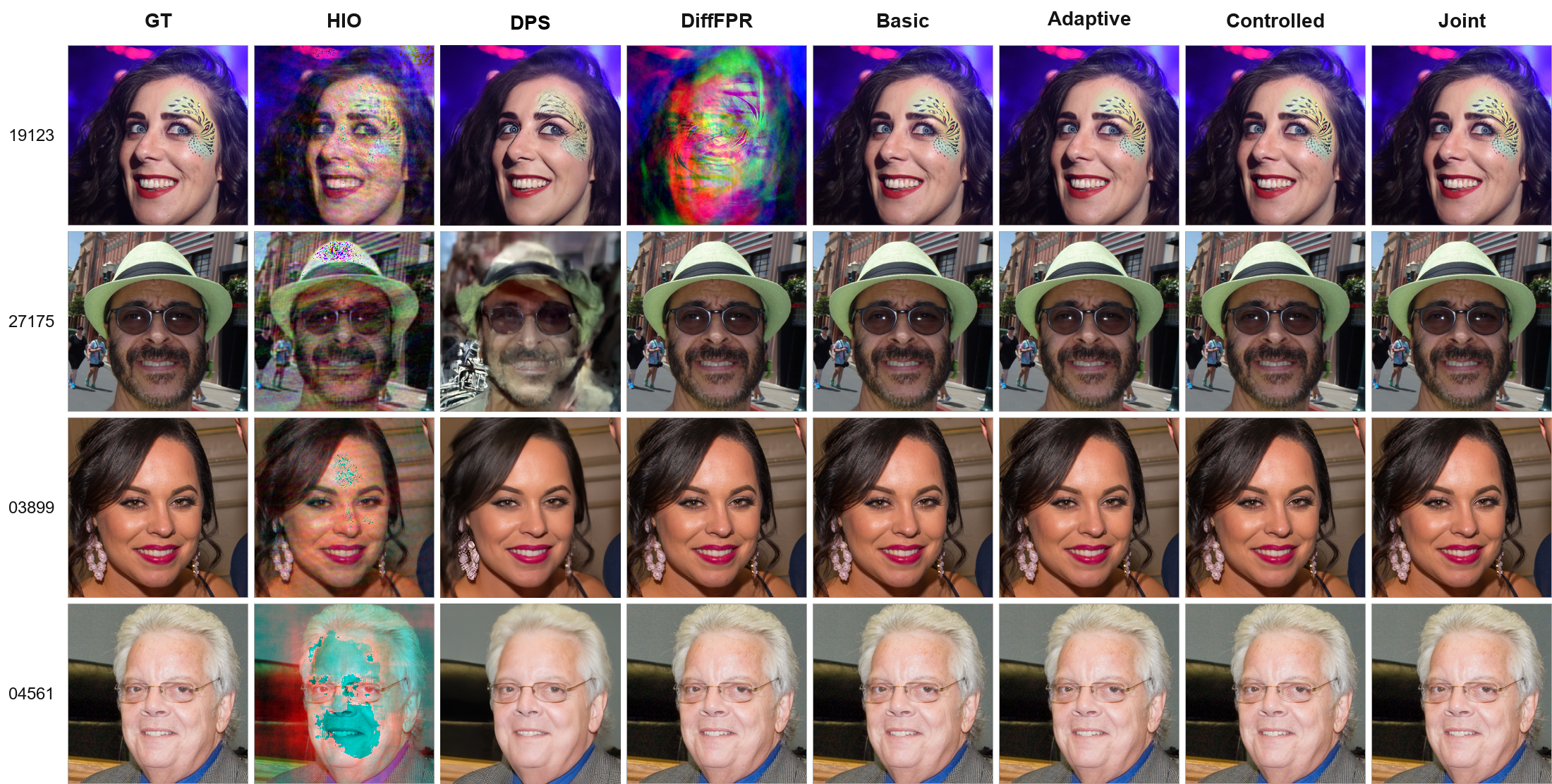}
\caption{Qualitative reconstructions under noiseless observations. Rows from top to bottom show images 19123, 27175, 03899, and 04561. Columns from left to right show ground truth, HIO, DPS, DiffFPR, Flow-RAAR-basic, Flow-RAAR-adaptive, Flow-RAAR-controlled, and the joint variant. DiffFPR follows the channelwise ambiguity-resolution rule in its public code.}
\label{fig:visual_noise000}
\end{figure*}

\begin{figure*}[!t]
\centering
\includegraphics[width=\textwidth]{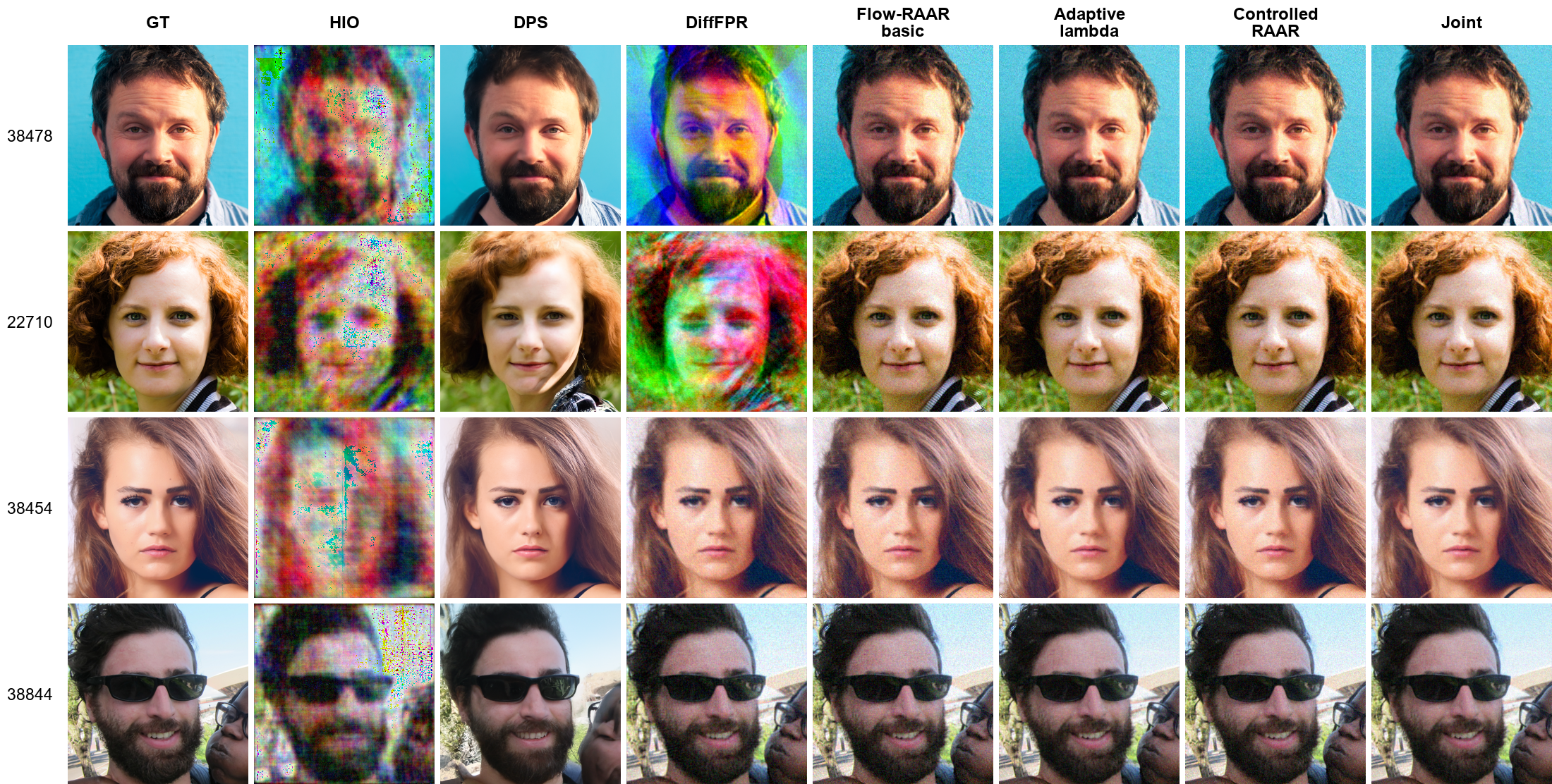}
\caption{Qualitative reconstructions at $\sigma=0.03$. Rows from top to bottom show images 38478, 22710, 38454, and 38844. The first two are randomly selected DiffFPR failures (PSNR below 20~dB), and the last two are successful cases (PSNR at least 20~dB). Method columns follow Fig.~\ref{fig:visual_noise000}.}
\label{fig:visual_noise003}
\end{figure*}

\begin{figure*}[!t]
\centering
\includegraphics[width=\textwidth]{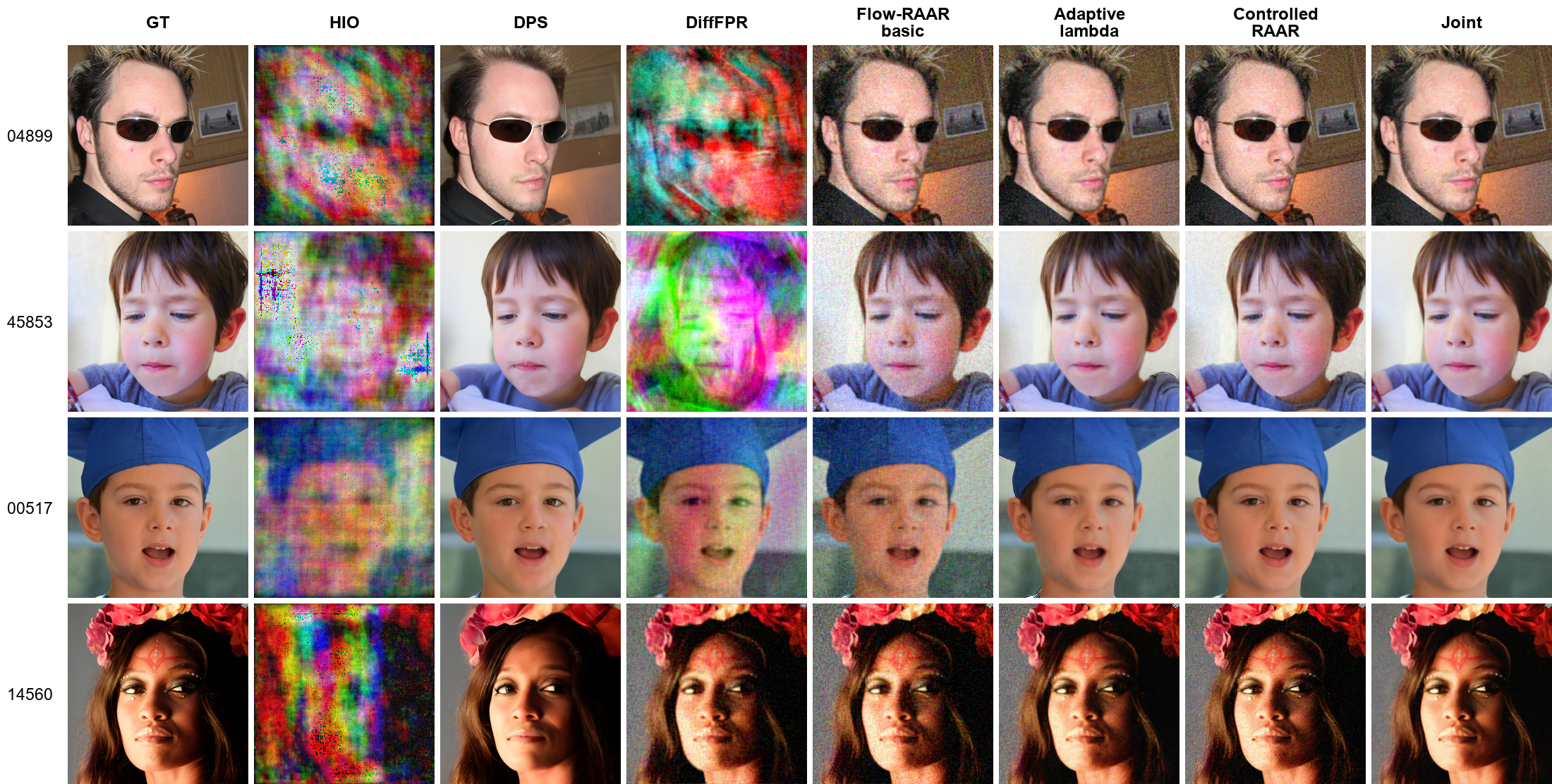}
\caption{Qualitative reconstructions at $\sigma=0.05$. Rows from top to bottom show images 04899, 45853, 00517, and 14560. The first two are randomly selected DiffFPR failures (PSNR below 20~dB), and the last two are successful cases (PSNR at least 20~dB). Method columns follow Fig.~\ref{fig:visual_noise000}.}
\label{fig:visual_noise005}
\end{figure*}

Figs.~\ref{fig:visual_noise000}--\ref{fig:visual_noise005} show representative reconstructions at the three noise levels. The Flow--RAAR methods recover image details well under noiseless observations; with noise, the residual-guided variants generally exhibit less noise and fewer artifacts, consistent with the quantitative trends in Table~\ref{tab:ffhq25_main_comparison}.

\subsection{Ablation Studies}
\label{subsec:ablation}

We do not use simple on--off combinations to ablate the two improvements because all four Flow--RAAR methods are already compared independently in Table~\ref{tab:ffhq25_main_comparison}. Instead, we examine the data-consistency operator, iteration depth, fusion schedule, and the control parameters and stopping behavior of Flow-RAAR-controlled. All ablations use the initial 25-image subset of the 50-image FFHQ test set, fourfold oversampling, and 20 random restarts. These are the first 25 identifiers in the separate appendix, not a quality-ranked subset. Their sample size remains 25; they are not pooled with the supplementary batch, so their means need not equal the 50-image means in Table~\ref{tab:ffhq25_main_comparison}. Apart from the factor under study and explicitly stated noise level, all parameters remain fixed.

\subsubsection{Flow-RAAR Versus Flow-HIO}
At $\sigma=0.03$, we retain the Flow prior, outer time interval, fusion schedule, outer iterations, and inner depth, replacing only the RAAR data-consistency operator with HIO. Table~\ref{tab:ablation_flow_hio} reports the results. Within the same generative-prior framework, Flow-RAAR-basic improves mean PSNR by 6.7021~dB and SSIM by 0.1631 and reduces LPIPS from 0.5608 to 0.2387. Its PSNR is higher for every one of the 25 images, and its relative magnitude residual decreases from 0.15853 to 0.11369. The difference therefore does not arise solely from the generative prior: RAAR's reflection and relaxed-averaging structure is better suited to alternating with Flow predictions, improving both consistency and image quality.

\begin{table}[t]
\caption{Ablation of RAAR and HIO Data-Consistency Operators Within the Flow-Prior Framework ($\sigma=0.03$).}
\label{tab:ablation_flow_hio}
\centering
\renewcommand{\arraystretch}{1.08}
\scriptsize
\setlength{\tabcolsep}{3.0pt}
\begin{tabular}{lcccc}
\toprule
Consistency operator & PSNR (dB) & SSIM & LPIPS & Relative residual \\
\midrule
HIO & 20.9394 & 0.4648 & 0.5608 & 0.15853 \\
RAAR & \textbf{27.6415} & \textbf{0.6279} & \textbf{0.2387} & \textbf{0.11369} \\
\bottomrule
\end{tabular}
\end{table}

\subsubsection{Outer and Inner Iteration Counts}
Table~\ref{tab:ablation_nk} reports results at $\sigma=0$ and $\sigma=0.03$. At each noise level, we vary $N\in\{400,800,1200\}$ with $K=10$ fixed and vary $K\in\{5,10,15\}$ with $N=800$ fixed. The product $NK$ is the number of RAAR updates per reconstruction trajectory and serves as an indicator of inner computational cost.

Without noise, increasing $N$ from 400 to 800 at $K=10$ raises PSNR from 46.1400 to 46.9896~dB, but increasing it to 1200 yields no further gain. With $N=800$, $K=15$ achieves 48.4409~dB, 0.9951 SSIM, and 0.0006 LPIPS, outperforming $K=5$ and $K=10$. More extensive inner magnitude correction can therefore improve accuracy with exact observations, while the gains from additional outer Flow iterations largely saturate around $N=800$.

\begin{table}[t]
\caption{Ablation of Outer Iterations $N$ and Fixed Inner Steps $K$ at Different Noise Levels.}
\label{tab:ablation_nk}
\centering
\renewcommand{\arraystretch}{1.08}
\scriptsize
\setlength{\tabcolsep}{3.5pt}
\begin{tabular}{cccrcccc}
\toprule
$\sigma$ & $N$ & $K$ & $NK$ & PSNR & SSIM & LPIPS & \shortstack{Relative\\residual} \\
\midrule
\multirow{6}{*}{$0$} & 400 & 10 & 4000 & 46.1400 & 0.9913 & 0.0012 & 0.00185 \\
 & 800 & 10 & 8000 & \textbf{46.9896} & \textbf{0.9934} & \textbf{0.0009} & \textbf{0.00175} \\
 & 1200 & 10 & 12000 & 46.9656 & 0.9930 & 0.0010 & 0.00176 \\
\cmidrule(lr){2-8}
 & 800 & 5 & 4000 & 43.9924 & 0.9872 & 0.0023 & 0.00272 \\
 & 800 & 10 & 8000 & 46.9896 & 0.9934 & 0.0009 & 0.00175 \\
 & 800 & 15 & 12000 & \textbf{48.4409} & \textbf{0.9951} & \textbf{0.0006} & \textbf{0.00145} \\
\midrule
\multirow{6}{*}{$0.03$} & 400 & 10 & 4000 & \textbf{27.7316} & \textbf{0.6314} & \textbf{0.2324} & 0.11372 \\
 & 800 & 10 & 8000 & 27.6415 & 0.6279 & 0.2387 & 0.11369 \\
 & 1200 & 10 & 12000 & 27.5913 & 0.6259 & 0.2409 & \textbf{0.11367} \\
\cmidrule(lr){2-8}
 & 800 & 5 & 4000 & \textbf{28.0051} & \textbf{0.6538} & \textbf{0.2042} & 0.11352 \\
 & 800 & 10 & 8000 & 27.6415 & 0.6279 & 0.2387 & 0.11369 \\
 & 800 & 15 & 12000 & 26.9722 & 0.6118 & 0.2612 & \textbf{0.11334} \\
\bottomrule
\end{tabular}
\end{table}

At $\sigma=0.03$, quality decreases as $N$ increases among the three settings with $K=10$. For $N=800$, $K=5$ achieves the highest PSNR and SSIM, 28.0051~dB and 0.6538, and lowest LPIPS, 0.2042; $K=15$ decreases PSNR to 26.9722~dB and increases LPIPS to 0.2612. Notably, $K=15$ has the lowest relative residual but the poorest image quality, indicating that increasingly enforcing noisy magnitudes can progressively overfit measurement noise.

The noise levels exhibit opposite trends: greater inner depth helps approach exact magnitude constraints without noise, while shallower settings are more robust with noise. Iteration depth must therefore balance consistency accuracy and noise fitting. We retain $N=800,K=10$ to use a common configuration across noise levels and Flow--RAAR variants, not to claim that it is optimal after separate parameter tuning at every noise level.

\subsubsection{Sensitivity to Fusion Schedules and Control Parameters}
The straight Flow path in \eqref{eq:flow_interpolation} assigns coefficients $t$ and $1-t$ to the current image and random component. At smaller $t$, the path state contains more randomness and the Flow prediction is more uncertain, motivating greater preservation of the generative prior and avoidance of excessive measurement correction. As $t$ increases, more current-image information is retained and RAAR correction can be strengthened. Inspired by the straight path's signal coefficient $t$, we use the linear baseline $g(t)=t$ without additional shape parameters, increasing the consistency weight linearly with time. This is a structurally natural baseline, but it is not uniquely determined by the straight path or necessarily optimal for other noise levels or probability paths.

To assess schedule shape, we compare $g(t)=\sqrt t$, $g(t)=t$, and $g(t)=t^2$ at $\sigma=0$ and $\sigma=0.05$. The sweep uses the same 25 FFHQ images from the main experiment, $\kappa=4$, $N=800$, $K=10$, and 20 random restarts. Since $\sqrt t>t>t^2$ for $t\in(0,1)$, these schedules correspond to aggressive, linear, and conservative early RAAR correction. Table~\ref{tab:sensitivity_schedule} reports the results.

\begin{table}[t]
\caption{Sensitivity to the Baseline Fusion Schedule $g(t)$ at Different Noise Levels.}
\label{tab:sensitivity_schedule}
\centering
\renewcommand{\arraystretch}{1.08}
\footnotesize
\setlength{\tabcolsep}{3.5pt}
\begin{tabular}{cccccc}
\toprule
$\sigma$ & $g(t)$ & PSNR (dB) & SSIM & LPIPS & \shortstack{Relative\\residual} \\
\midrule
\multirow{3}{*}{$0$} & $\sqrt t$ & \textbf{47.0630} & 0.9930 & \textbf{0.00090} & \textbf{0.00158} \\
& $t$ & 46.9896 & \textbf{0.9934} & 0.00095 & 0.00175 \\
& $t^2$ & 45.3022 & 0.9890 & 0.00490 & 0.00281 \\
\midrule
\multirow{3}{*}{$0.05$} & $\sqrt t$ & 24.1720 & 0.4541 & 0.4357 & 0.19266 \\
& $t$ & 24.2539 & 0.4587 & 0.4303 & 0.19255 \\
& $t^2$ & \textbf{24.3626} & \textbf{0.4661} & \textbf{0.4243} & \textbf{0.19233} \\
\bottomrule
\end{tabular}
\end{table}

Without noise, $\sqrt t$ achieves the highest PSNR and lowest LPIPS and residual, while $t$ achieves the highest SSIM. Their PSNR differs by only 0.0734~dB, with LPIPS values 0.00090 and 0.00095. The overly conservative $t^2$ reduces PSNR to 45.3022~dB. Under strong noise, the trend changes: $t^2$ suppresses early correction toward noisy measurements and slightly improves all three quality metrics over the linear schedule, while $\sqrt t$ performs worst. The preferred shape thus depends on noise level. We retain $g(t)=t$, motivated by the straight path's signal coefficient without additional shape parameters, as the common baseline and leave noise-adaptive nonlinear schedules for future optimization.

Early stopping uses the threshold $\rho\hat r_n$. Table~\ref{tab:sensitivity_rho} reports threshold sensitivity at $\sigma=0$ and $\sigma=0.05$, using the same initial 25 FFHQ images and $K_{\max}=10$. Without noise, $\rho=0.85$ achieves the highest PSNR and SSIM and the lowest LPIPS in this sweep. Increasing $\rho$ reduces the mean number of inner iterations performed, $\bar J$, from 1.497 to 1.039, but degrades reconstruction quality. With noise, $\rho=0.90$ achieves the best values of all three quality metrics, whereas $\rho=0.95$ yields slightly lower quality, indicating that the preferred threshold depends on the noise conditions. Considering both noiseless reconstruction accuracy and the use of a common configuration across noise levels, we select $\rho=0.85$ for the main experiments rather than tuning it separately at each noise level. The table distinguishes the mean number of inner iterations performed, $\bar J$, from the mean selected candidate index, $\bar j^\star$.

\begin{table}[t]
\caption{Sensitivity to the Flow-RAAR-controlled Threshold $\rho$ at Different Noise Levels. Here, $\bar J$ is the mean number of inner iterations performed, and $\bar j^\star$ is the mean selected candidate index.}
\label{tab:sensitivity_rho}
\centering
\renewcommand{\arraystretch}{1.08}
\scriptsize
\setlength{\tabcolsep}{3.5pt}
\begin{tabular}{ccccccc}
\toprule
$\sigma$ & $\rho$ & PSNR (dB) & SSIM & LPIPS & $\bar J$ & $\bar j^\star$ \\
\midrule
\multirow{3}{*}{$0$} & 0.85 & \textbf{42.2200} & \textbf{0.9762} & \textbf{0.0070} & 1.497 & 1.497 \\
 & 0.90 & 40.4453 & 0.9685 & 0.0119 & 1.210 & 1.210 \\
 & 0.95 & 38.5404 & 0.9604 & 0.0166 & \textbf{1.039} & \textbf{1.039} \\
\midrule
\multirow{3}{*}{$0.05$} & 0.85 & 27.2900 & 0.6187 & 0.2156 & 8.155& 6.673 \\
 & 0.90 & \textbf{27.6565} & \textbf{0.6354} & \textbf{0.1990} & 6.668 & 5.301 \\
 & 0.95 & 27.6370 & 0.6339 & 0.2026 & \textbf{3.822} & \textbf{2.750} \\
\bottomrule
\end{tabular}
\end{table}

\subsubsection{Ablation of Flow-RAAR-controlled Early Stopping}
At $\sigma=0.03$ and $K_{\max}=10$, we compare the main method with early stopping in \eqref{eq:controlled_stop} against a setting that disables early stopping and always runs to $K_{\max}$. Both retain best-so-far selection, isolating the stopping rule's effect. Controller statistics follow the main restart-selection procedure: first select the best of 20 restarts for each image, then compute mean number of inner iterations performed and mean selected candidate index along that trajectory, and finally average over the 25 images.

Disabling early stopping fixes $J_n=K_{\max}$. Let $j_n^\star$ denote the best-so-far candidate index. For one reconstruction trajectory, we first average over its $N$ outer iterations:
\begin{equation}
    \bar J=\frac{1}{N}\sum_{n=0}^{N-1}J_n,
    \qquad \bar j^\star=\frac{1}{N}\sum_{n=0}^{N-1}j_n^\star.
\label{eq:controller_statistics}
\end{equation}
The table further averages $\bar J$ and $\bar j^\star$ over the 25 images. The former measures the number of RAAR iterations performed; the latter measures the mean inner iteration index of the selected candidate. They are distinct quantities.

Table~\ref{tab:ablation_early_stop} shows that early stopping with $K_{\max}=10$ reduces mean number of inner iterations performed from 10 to 5.047, eliminating approximately 49.5\% of RAAR updates, while improving PSNR by 0.3701~dB and reducing LPIPS from 0.1797 to 0.1731. The setting without early stopping still returns the best-so-far candidate over the full path, so merely computing additional candidates does not directly change the state used for fusion. Instead, expanding the candidate set from $\{0,\ldots,J_n\}$ to $\{0,\ldots,K_{\max}\}$ admits deeper candidates that have lower residuals but may fit measurement noise more closely and replace shallower candidates. Early stopping therefore reduces computation and truncates the candidate search space, reducing the risk of selecting a candidate that overfits measurement noise for fusion. The mean selected candidate index without early stopping is 9.577 rather than exactly 10, confirming that best-so-far selection remains active.

\begin{table}[t]
\caption{Ablation of the Flow-RAAR-controlled Early Stopping Rule ($\sigma=0.03$, $K_{\max}=10$).}
\label{tab:ablation_early_stop}
\centering
\renewcommand{\arraystretch}{1.08}
\setlength{\tabcolsep}{3.5pt}
\begin{tabular}{lccccc}
\toprule
Stopping rule & PSNR & SSIM & LPIPS & $\bar J$ & $\bar j^\star$ \\
\midrule
Run to $K_{\max}$ & 28.3409 & 0.6758 & 0.1797 & 10.000 & 9.577 \\
Early stopping & \textbf{28.7110} & \textbf{0.6777} & \textbf{0.1731} & \textbf{5.047} & \textbf{4.654} \\
\bottomrule
\end{tabular}
\end{table}

\subsection{Runtime Comparison}
\label{subsec:runtime_comparison}

Following Section~\ref{subsec:experimental_setup}, we compare DPS, DiffFPR, and our methods at $\sigma=0.05$ using the same 25 FFHQ images, $256\times256$ image size, oversampling parameter $\kappa=4$, and 20 random restarts per image. All methods run on the same NVIDIA GeForce RTX 4070 SUPER GPU. Timing includes the end-to-end wall-clock time for reconstructing all 25 images, completing 20 restarts per image, and selecting the best results. The mean in Table~\ref{tab:runtime_comparison} is the total divided by 25 and thus measures the average cost of 20 restarts and selection for one image. Speedup is normalized to DiffFPR at $1.00\times$; values above one indicate faster execution.

\begin{table}[t]
\caption{Runtime Comparison on 25 Images Using the Same RTX 4070 SUPER, where DiffFPR is selected as the baseline.}
\label{tab:runtime_comparison}
\centering
\renewcommand{\arraystretch}{1.08}
\scriptsize
\setlength{\tabcolsep}{3pt}
\begin{tabular}{lccc}
\toprule
Method & \shortstack{Total time\\(h)} & \shortstack{Mean time\\(s/image)} & \shortstack{Relative\\speedup} \\
\midrule
DiffFPR & 5.229 & 753.04 & 1.00 \\
DPS & 8.953 & 1289.24 & 0.58 \\
Flow-RAAR-basic & 3.463 & 498.60 & 1.51 \\
Flow-RAAR-adaptive & 4.508 & 649.08 & 1.16 \\
Flow-RAAR-controlled & \textbf{3.410} & \textbf{491.08} & \textbf{1.53} \\
Flow-RAAR-joint  & 3.812 & 548.91 & 1.37 \\
\bottomrule
\end{tabular}
\end{table}

DPS and DiffFPR perform 1000 reverse-diffusion steps per trajectory. DPS additionally backpropagates the magnitude-error gradient to the current diffusion state at each step, requiring 1289.24~s per image, 71.2\% longer than DiffFPR. All four Flow--RAAR methods use 800 outer Flow updates. The basic method reduces mean end-to-end time from 753.04~s for DiffFPR to 498.60~s, a 33.8\% reduction and $1.51\times$ speedup. Flow-RAAR-adaptive requires 649.08~s per image and achieves $1.16\times$ speedup. Its timing includes residual evaluation, diagnostic recording, and CPU--GPU synchronization, so the increase over the basic method cannot be attributed entirely to residual evaluation. Flow-RAAR-controlled is the fastest Flow variant at 491.08~s per image, a 34.8\% reduction and $1.53\times$ speedup over DiffFPR. The joint variant requires 548.91~s, a 27.1\% reduction and $1.37\times$ speedup. Under common hardware and experimental scale, all four Flow--RAAR methods are faster than DPS and DiffFPR.

\section{Conclusion}
\label{sec:conclusion}

We presented FlowFPR for oversampled Fourier phase retrieval of natural RGB images. The framework couples a time-dependent Flow prediction with RAAR-type measurement correction and reconstructs the next path state after candidate fusion. Residual-guided adaptive fusion and controlled inner search yield two independent improvements and a joint variant. Under the ground-truth-assisted best-of-20 protocol on 50 FFHQ images, the joint variant attains the best mean PSNR, SSIM, and LPIPS among the eight evaluated methods at both tested nonzero noise levels, whereas the basic method performs best without noise. Runtime measurements on the same hardware on the initial 25-image subset shows lower end-to-end runtime for all four FlowFPR instances than for DPS and DiffFPR.

These findings are limited to the tested face images, magnitude-domain Gaussian noise, and oracle restart-selection protocol; they do not establish superiority for a single run or under arbitrary noise conditions. Future work will examine ground-truth-free restart selection, broader image distributions, and realistic measurement noise, together with uncertainty and runtime analyses based on larger-scale experiments.

\appendices
\section{FFHQ Test Image Identifiers}
\label{app:ffhq_image_ids}
The following identifiers list all 50 FFHQ test images in experimental batch order, retaining leading zeros. The initial batch was used for ablations and timing; the supplementary batch was sampled after excluding the initial images.

\subsection{Initial Batch (25 Images)}
47730, 53427, 19123, 42810, 08886, 00528, 37770, 21380, 32402, 18479, 12957, 08248, 03899, 22372, 04029, 46668, 38454, 14560, 38932, 32774, 16467, 27175, 04561, 50548, 19453.

\subsection{Supplementary Batch (25 Images)}
28665, 00517, 53739, 03306, 00093, 02828, 38844, 51866, 36825, 45853, 04501, 16596, 27826, 03292, 13201, 17078, 38478, 38705, 17279, 04899, 41822, 37844, 45008, 22710, 30685.

\bibliographystyle{IEEEtran}
\bibliography{references}

\end{document}